\documentstyle[eqsecnum,epsfig,amssymb,aps,prb,preprint]{revtex}

\tightenlines

\begin{document}
\author{Dirk Reith\footnote{corresponding author} and Florian M\"uller-Plathe}
\address{
  Max-Planck-Institut f\"ur Polymerforschung,
  Ackermannweg 10, D-55121 Mainz, Germany}
\title{On the nature of Thermal Diffusion in binary Lennard-Jones liquids}
\date{\today}
\maketitle

\abstract{The aim of this study is to understand deeper 
  the thermal diffusion transport process (Ludwig-Soret effect) at the 
  microscopic level. For that purpose, the recently developed reverse 
  nonequilibrium molecular dynamics method was used to calculate 
  Soret coefficients of various systems in a systematic fashion. We studied
  binary Lennard-Jones (LJ) fluids near the triple point (of one of the
  components) in which we separately changed the ratio of one of the 
  LJ parameters mass, atomic diameter and interaction strength while 
  keeping all other parameters fixed and identical.
  We observed that the magnitude of the Soret coefficient depends on all 
  three ratios. Concerning its sign we found that heavier species, 
  smaller species and species with higher interaction strengths tend to 
  accumulate in the cold region whereas the other ones (lighter, bigger or 
  weaker bound) migrate to the hot region of our simulation cell.
  Additionally, the superposition of the influence of the various parameters 
  was investigated as well as more realistic mixtures. We found that in the 
  experimentally relevant parameter range the contributions are nearly 
  additive and that the mass ratio often is the dominating factor. }
\vfill

\section{INTRODUCTION}
Transport processes play an important role in the understanding of the
properties of liquid mixtures. Thermal diffusion (the Ludwig-Soret effect) is 
one of these intriguing processes. It characterizes the flux of matter in response 
to a gradient in temperature and has already been studied for almost 150 years:
The migration of atoms and molecules as a consequence of a temperature gradient 
was first reported by Ludwig~\cite{ludwig:1856} when studying sodium-sulfate solutions
in 1856. Later, in 1879, Soret~\cite{soret:1879} observed the same effect in other
electrolyte solutions. In the case of gases, it was predicted independently
by Enskog~\cite{enskog:12} and by Chapman~\cite{chapman:17a}, and later 
confirmed by the experiments of Chapman and Dootson~\cite{chapman:17b}. 
Further, several small organic molecule or polymer mixtures have been 
investigated by various experimental methods such as thermo-gravitational 
columns (e.g.\ Clusius and Dickel~\cite{clusius:38}), thermal
field flow fractionation (e.g.\ Giddings~\cite{giddings:76}) or most recently
by thermal diffusion forced Rayleigh scattering 
(K\"{o}hler~\cite{koehler:95b}). There are 
numerous examples for the technological significance of thermal 
diffusion, e.g.\ presented by Kincaid, Ratkje and 
Hafskj{\o}ld~\cite{kincaid:94,hafskjold:96b}. 
However, there is still no satisfactory theory to explain the effect.
``It is the only hydrodynamic transport mechanism that lacks a simple physical
explanation.''~\cite{kincaid:94}. But the situation is even worse:
One cannot even approximately predict the transport coefficients of closely related 
systems. The size of the Ludwig-Soret effect differs e.g.\ for equimolar mixtures of 
benzene-chlorobenzene and benzene-nitrobenzene roughly by a factor 
of 10~\cite{d'ans:92}. Additionally, the experimental data base is rather small and 
partly inconsistent~\cite{koehler:95b}. One reason for this is probably that 
thermally driven flows of matter are minor effects and thus 
several orders of magnitude smaller than concentration driven flows.

In multi-component systems, couplings between the different types of transport
are, on a fundamental level, described by Onsager's linear 
relationships~\cite{onsager:31}:

\begin{eqnarray}  \label{eq01}
\vec{J}_{k} &=&\sum_{j=1}^{N}L_{kj}\vec{X}_{j}+L_{kq}\vec{X}_{q}\hspace{30pt}%
k=1,...,N \\
\vec{J}_{q} &=&\sum_{j=1}^{N}L_{qj}\vec{X}_{j}+L_{qq}\vec{X}_{q}  \nonumber
\end{eqnarray}

with $\vec{J}_{k}$ being the flux of component $k$ in 
$\mbox{gcm}^{-2}\mbox{s}^{-1}$ with
respect to the center of mass of the system and $\vec{J}_{q}$ being the heat
flux in $\mbox{Jcm}^{-2}\mbox{s}^{-1}$. The $\vec{X}_{\alpha }$ are 
thermodynamic forces,
specifically for a binary mixture $\vec{X}_{q}=-\frac{\vec{\nabla}T}{T}$ and 
$\vec{X}_{j}=-\vec{\nabla}_{T}(\mu _{1}-\mu_{2})$. (The
subscript $T$ denotes that the gradient has to be taken under isothermal
conditions.) The $L_{\alpha \beta }$ are the 
Onsager
coefficients with $\alpha=(k,q)$ and $N$ is number of species in the mixture.
Note that this representation assumes $\vec{J}%
\Vert \vec{X} $, otherwise each $L_{\alpha \beta }$ has to be a 
$3\times 3$ tensor. The assumption of linear response holds
for many transport processes through condensed media. In this picture, thermal
diffusion is characterized by the matrix coefficient $L_{kq}$.

Although equations~\ref{eq01} have been known for several decades, there is yet 
no deeper insight into the microscopic mechanisms of the \textsl{coupling} of 
heat and mass transport. Liquid mixtures are often difficult to understand by 
analytical theories. Performing computer simulations, in contrast, is a way to look 
easily at the microscopic properties of a liquid system. The most 
promising results to understand thermal diffusion seem to come from molecular 
dynamics (MD) simulations. Several MD techniques have been developed in order to 
study the Ludwig-Soret effect. The early algorithms suffered from conceptual 
as well as practical problems: firstly, of how to define and compute the heat flow 
or the heat of transfer accurately in the microscopic picture and, secondly, from 
the large perturbation fields that were necessary to observe thermal 
diffusion~\cite{macgowan:86,macgowan:87,paolini:87,vogelsang:88,evans:87}).
Hafskj{\o}ld et al.\ ~\cite{hafskjold:94} were the first to develop an 
algorithm that side-steps these problems. In fact, their methods are similar 
in spirit to ours. They define microscopic fluxes in an unambiguous way. However, 
they mostly investigated dilute gas mixtures~\cite{kincaid:94} or 
interfacial systems~\cite{hafskjold:94,hafskjold:96a,hafskjold:97} so far. 
To our knowledge, there is no systematic study of thermal diffusion in the 
high density liquid regime up to present. 

In this contribution, we study both qualitative and quantitative 
aspects of the Ludwig-Soret effect. This is done by performing 
reverse non-equilibrium molecular dynamics (RNEMD) computer simulations
for binary Lennard-Jones (LJ) systems. We investigate the influence of
systematic variations of the physical
parameters (mass $m$, atomic diameter $\sigma$ and interaction strength
$\varepsilon$) of our model system: By changing one parameter ratio 
(e.g.\ $m_{1}/m_{2}$) while keeping all other parameters fixed and identical, we 
are able to observe exactly how the Ludwig-Soret effect depends on every one 
of them separately. We exploit the advantage of better control of the system
variables in computer simulation over experiment to better understand thermal 
diffusion. We discuss our simulation results in the framework of linear irreversible 
thermodynamics. Moreover, we establish empirical rules for the influence of 
differences in the individual molecular parameters on the Ludwig-Soret effect 
and finally investigate to which extent they are additive.

\section{THEORY}
\subsection*{Linear Irreversible Thermodynamics \label{irrTD}}
In order to define proper quantities in the theory of linear irreversible 
thermodynamics, the 
assumption of local equilibrium is essential. It enables us to apply the 
well-known equilibrium machinery to local volume elements of a 
perturbed system by defining intensive quantities derived from the extensive 
equilibrium quantities~\cite{evans:90}.
The entropy source strength $s$  given in $\frac{J}{s\cdot K\cdot cm^{3}}$
is the fundamental quantity of this theory. Due to Onsager's  
variational principle~\cite{onsager:31}, $s \geq 0 $  
has to be minimized. The non-equilibrium steady state is the one in which 
the least amount of total entropy is produced:
\begin{equation}
  \int_{V}s(\vec{r},t) dV \quad\longrightarrow\quad\mbox{min!}\nonumber
\end{equation}
It is also the state in which heat is most 
efficiently conducted through the system. Using a local form of the Gibbs 
equation in 
connection with the continuity equation for the entropy (for details, see 
e.g.\  ~\cite{drdip:98}) leads to the following equation for $s$ in a 
binary system with coupled flux of heat and matter:
\begin{equation}
s(\vec{r},t) =-\left[\vec{J}_{q}(\vec{r},t)\cdot\vec{\nabla}T(\vec{r},t)\right]%
\frac{1}{T(\vec{r},t)^{2}}-\left[\vec{J}_{1}\cdot\vec{\nabla}_{T}\mu(\vec{r},t)%
\right]\frac{1}{T(\vec{r},t)} 
\label{eq2}
\end{equation}
with $\vec{J}_{1}$ being the mass flux of species 1 in $\frac{g}{cm^{2}s}$ with
respect to the center of mass of the system, $\vec{J}_{q}$ the heat flux in $%
\frac{J}{cm^{2}s}$, $\mu = \mu_{1} - \mu_{2} $ the effective chemical 
potential in $\frac{J}{g}$ and $T$ the temperature in $K$. The right hand 
side of 
equation~\ref{eq2} is the sum of products of the fluxes (of heat and matter), 
thermodynamic forces (the gradients) and of 'thermodynamic factors' 
$\frac{1}{T^{2}}$ and $\frac{1}{T}$, respectively. The latter provide for 
the correct dimensions. In this article, we consider the time-independent 
steady state in which the matter flux has died out and a
constant temperature gradient is the only origin for the entropy source 
strength:
\begin{equation}
  s(\vec{r})_{\vec{J}_{1}=0} =-\left[\vec{J}_{q}(\vec{r})%
  \cdot\vec{\nabla}T\right]\frac{1}{T(\vec{r})^{2}} .
  \label{eq3}
\end{equation}
Combining this with Fourier's law for heat conduction we get: 
\begin{equation}
  s(\vec{r})_{\vec{J}_{1}=0} =-\lambda_{av}\left(\frac{\vec{\nabla T}}%
  {T(\vec{r})}\right)^{2} .
  \label{eq4}
\end{equation}

The thermal conductivity depends on density, temperature and concentration of
the mixture, which may vary over the system. Still, we assume that 
the perturbation and, hence, the variation in these quantities is small
enough, so it is sensible to use an average thermal conductivity $\lambda_{av}$.
The magnitude of the temperature gradient $\nabla T$ is the decisive 
quantity of the entropy source strength in our binary system. Note finally, 
that the local entropy production increases towards the cold region of 
the system as $1/T(\vec{r})^{2}$.
\subsection*{Phenomenological Transport Coefficients}
Equations~\ref{eq01} give the impression of a universal, clean and symmetric 
($L_{\alpha\beta}=L_{\beta\alpha}$) theory. Unfortunately, they have the 
disadvantage that the quantities of interest (i.e.\ the Onsager coefficients 
and the thermodynamic forces) are related to but not identical with 
experimentally measurable transport coefficients and fields. The
relations between them and the Onsager coefficients $L_{\alpha\beta}$ 
are well known (cf.\ e.g.\ ~\cite{hafskjold:93,drdip:98}). In their analysis, 
RNEMD simulations are akin to experiment~\cite{mueller-plathe:99}, 
so in the following we will concentrate on the relevant experimental 
transport coefficients:  the self diffusion coefficients $D_{k}$ $(k=1,2)$ 
(dimension $cm^{2}/s$) and the Soret coefficient $S_{T}$ (dimension $1/K$).
The former are calculated in our simulations via the Einstein route:
\begin{equation} \label{eq05}
D_{k}=\frac{1}{6}\frac{d}{dt}\left\langle \left| \vec{r}_{j}^{(k)}(t)-\vec{r}%
_{j}^{(k)}(0)\right| ^{2}\right\rangle .
\end{equation}
Averaging is performed over time origins as well as over all particles $j$ of 
type $k$. Since we empirically found that the $D_{k}$ calculated out of RNEMD 
simulations are (within statistical uncertainty) identical with 
equilibrium data (as we checked for some systems), we skipped further
equilibrium runs to save resources and use the RNEMD data. 
The Soret coefficient is defined as $S_{T}=D_{T}/D_{12}$ and has the 
physical meaning of the relative strength between
thermally induced diffusion (characterized by the thermal diffusion 
coefficient $D_{T}$) and concentration-driven, Fickian
diffusion (characterized by the mutual diffusion coefficient $D_{12}$). 
Phenomenologically, thermal diffusion is often expressed as
\begin{equation} \label{eq06}
  J_{1}= -D_{12}\rho \left[ \left(\frac{\partial w_{1}}{\partial z}\right)
   +S_{T}w_{1} (1-w_{1})\left(\frac{\partial T}{\partial z}\right)\right] .
\end{equation}
Here, we are assuming field (temperature gradient) and fluxes (energy and
matter) in z direction.
$J_{1}$ is the flux of species 1, $\rho$ the average mass density 
(assuming that the temperature gradient and the resulting density gradient
are small) and $w_{1}=x_{1}m_{1}/(x_{1}m_{1}+x_{2}m_{2})$ the weight fraction 
of species 1 ($x_{k}$ denotes the mole fraction of species $k$). 
Equation~\ref{eq06} appears to be similar to Onsager's description 
(Equations~\ref{eq01}). Note, however, that the gradient of the chemical 
potential has been replaced by the experimentally more accessible and
technologically more relevant weight fraction gradient. If the system is
continuously subjected to a temperature gradient, it will be driven to a 
non-equilibrium steady state: energy (heat) is then constantly flowing 
through it while the mass flux has stopped ($J_{1}=0$) and a constant
concentration (or weight fraction) gradient has been established.
Equation~\ref{eq06} can then be simplified and $S_{T}$, after application 
of the chain rule,  be obtained from the
temperature and molar fraction gradients in the system:
\begin{equation} \label{eq07}
  S_{T}=-\frac{1}{w_{1}\left( 1-w_{1}\right) }\left( \frac{\partial w_{1}}{%
      \partial z}\right) \left( \frac{\partial T}{\partial z}\right) ^{-1}  
=-\frac{1}{x_{1}\left( 1-x_{1}\right) }\left( \frac{\partial x_{1}}{%
      \partial z}\right) \left( \frac{\partial T}{\partial z}\right) ^{-1}.
\end{equation}
In the special case of equimolar mixtures ($x_{1}=x_{2}=0.5$),
equation~\ref{eq07} reduces further to 
\begin{eqnarray} \label{eq08}
  S_{T}=-4 \left( \frac{\partial x_{1}}{%
      \partial z}\right) \left( \frac{\partial T}{\partial z}\right) ^{-1}
\end{eqnarray}
The gradients appearing in equation~\ref{eq08}
can be calculated directly in our RNEMD simulations, 
making the analysis of the simulation data swift and easy. In particular,
$S_{T}$ can be calculated without previous evaluation of $D_{T}$ and $D_{12}$. 
Throughout this work, positive values of $S_{T}$ signify, that species 1 tends 
to accumulate in the \textsl{cold} regions of the simulation box.
\section{COMPUTATIONAL DETAILS}
\subsection*{Model System and Simulation Details}
All simulations are performed with 1500 Lennard-Jones (LJ) 
atoms~\cite{allen:87}
with cutoff distance $r_{c}=2.5\sigma_{1}$,  $\sigma_{1}\geq\sigma_{2}$. The
potential is defined as: 
\begin{equation}
U_{LJ}\left( r \right) =4\varepsilon \left[ \left( \frac{\sigma }{r}%
\right) ^{12}-\left( \frac{\sigma }{r}\right) ^{6}\right].  \label{LJpot}
\end{equation}

We investigate equimolar mixtures of 
two species only. Consequently, our model system has  six physical parameters:
atomic masses $m_{k}$ $(k=1,2)$, atomic diameters (in form of $\sigma_{k}$) 
and the interaction strengths (in form of $\varepsilon_{k}$) of our two model
species. For the interaction between unlike particles, the
Lorentz-Berthelot mixing rules~\cite{allen:87} are applied. No corrections to 
the interaction parameters due to excess functions (e.g.\ molar volume and
enthalpy) are considered, although
that can be done and might be useful in case of more realistic simulations. 
The orthorhombic periodic simulation cell is of size 
($L^{*}$ x $L^{*}$ x $3L^{*}$), with $L^{*}$ being roughly 
$8-13$. The asterisk indicates Lennard-Jones reduced units~\cite{allen:87}, 
using LJ-Argon values as reference values 
(Table~\ref{reduced_scales_table}). The volume is held constant during 
the simulation. Our simulations are performed  in the dense liquid
state, i.e.\ $\bar{\rho}^{*}=0.8-0.85$ and $\bar{T}^{*}=0.75-0.85$ 
(Table~\ref{reduced_scales_table}), which is close to the triple point 
of species 2. 
The average temperature $\bar{T}^{*}$ is maintained at its value by the 
weak-coupling scheme of Berendsen~\cite{berendsen:84} with a coupling time 
of $t_{coup}^{*} = 4.66$. We do so to prevent long-time drifts which might 
occur due to limited precision of the discrete, stepwise
solution of the equations of motion. They are integrated using the velocity-Verlet
algorithm in connection with a multiple-time-step
scheme~\cite{tuckerman:91} for the force calculations. 
Long-range forces ($r > 1.7\sigma _{m}$ with $\sigma _{m}=max\{\sigma
_{1},\sigma _{2}\}$) are evaluated only every 4 time steps
$\Delta t^{*}=4.66\cdot 10^{-3}$, the switching range is $0.1\sigma _{m}$.
Our simulations are performed with run lengths of $3\cdot 10^{6} - 10\cdot 10^{6}$ 
steps. Shorter preparatory-runs of $0.8\cdot 10^{6}-1.5\cdot 10^{6}$ steps are made 
to establish the non-equilibrium steady state. The runs have to be that long because
the fluctuations in the mole fraction profile are relatively large and 
thermal diffusion is a weak effect compared to 
other transport properties. Positions and velocities are written
out every $500-3000$ steps, local temperature and concentration in the slabs 
(see below) every $50-100$ steps for calculation of the relevant gradients.

\subsection*{The RNEMD algorithm}
\begin{figure}[hb]
\begin{center}
\epsfxsize 12cm \epsfbox {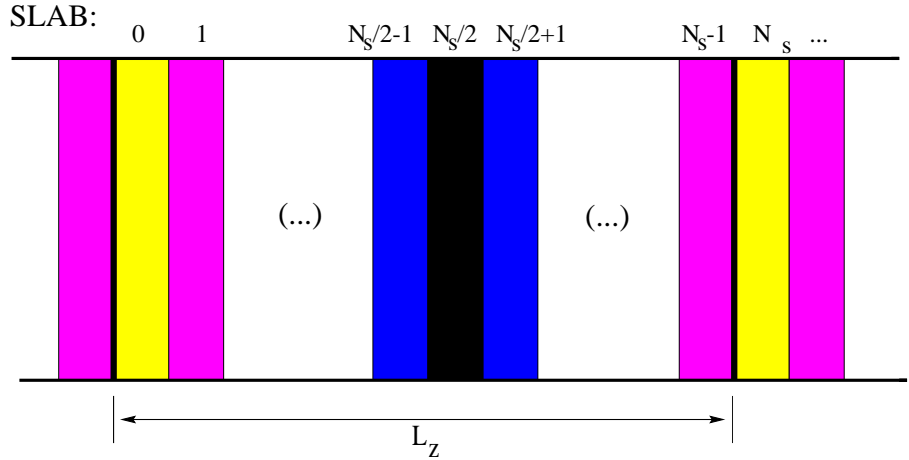}
\end{center}
\caption{Subdividing the periodic simulation box into slabs. Slab $0$ is 
  defined to be the cool slab, slab $N_{s}/2$ to be the hot slab. Kinetic 
  energy is
  artificially transferred by the heat exchange algorithm and flows back by
  thermal conduction.}
\label{F_box}
\end{figure}

\begin{figure}[hb]
\begin{center}
\epsfxsize 12cm \epsfbox {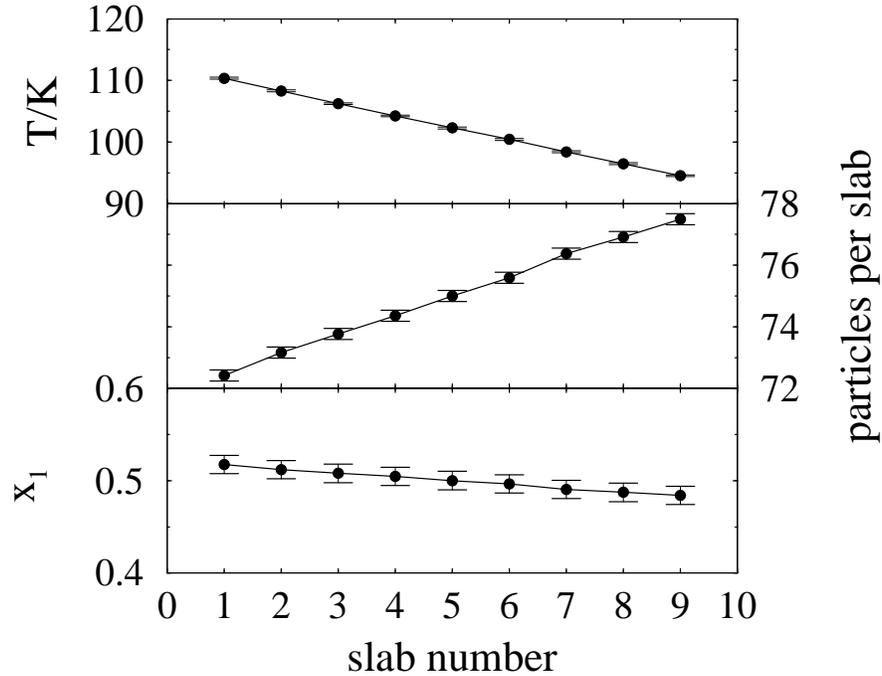}
\end{center}
\caption{Typical profiles of system properties at the state point
  $\bar{T}^{*}=0.81$, $\bar{\rho}^{*}=0.85$. $\sigma_{1} / \sigma_{2}=1.2$ 
  and species 2 corresponds to Lennard-Jones Argon. Slab $0$ is defined to be the 
  hot slab, slab $10$ to be the cool slab.}
\label{Fmix_example}
\end{figure}

The RNEMD method reverses the usual cause-and-effect picture of non
equilibrium simulations. For the study of thermal diffusion the 
``effect'', the heat flux, is imposed on the system whereas
the ``cause'', the temperature gradient is obtained from the simulation:
We divide the simulation cell into $N_{s}=20$ slabs perpendicular to the 
$z$-direction (Fig.~\ref{F_box}). The slabs are
chosen to be equally thick, i.e.\ to have identical volumes, and large enough
so that reasonable statistics can be expected: each slab contains 75
particles on average. A flow of heat $\vec{J}_{q}$ is artificially 
maintained by exchanging velocities of suitably selected particles : 
Slab $0$ is defined as the
'cool' slab and slab $N_{s}/2$ as the 'hot' slab. Every $N_{exch}$ steps, we
search through all atoms of the cold slab and determine the hottest ones of
both species. In the hot slab, we proceed conversely and determine the
coldest atoms of both species. Then we exchange
the velocities of the so determined atoms of the same species. We
were always able to find at least one particle in the
cold slab that is hotter than the coldest particle of the hot slab, 
since the Maxwell-Boltzmann distribution of atomic
kinetic energies is very broad compared to the temperature difference of the
two slabs. The exchange period has to be adjusted such that 
the perturbation is weak enough for linear response to hold (typically: 
$N_{exch} = 50 - 200$ steps). This leads to a temperature variation of
$ 2(T^{*}_{hot}-T^{*}_{cold}) / (T^{*}_{hot}+T^{*}_{cold}) \approx 0.1-0.15$.
It was shown~\cite{drdip:98,mueller-plathe:99,mueller-plathe:97a} that for
temperature differences of up to half the average temperature linear response
is still approximately fulfilled. The unphysical velocity exchange leads to 
a \textsl{physical} flux of heat in the opposite direction through 
the intervening slabs. As the response is linear, the procedure causes
linear profiles of temperature, overall density and concentration 
(Fig.~\ref{Fmix_example}). At steady state, physical and unphysical heat 
flux have the same magnitude because of energy conservation. Thus, cf.\ 
equation~\ref{eq4}, the temperature gradient will be minimal. To avoid
a possible breakdown of the local equilibrium due to the unphysical energy 
transfer in the thermostatting slabs, these are excluded from gradient calculations. 
General properties and details of the RNEMD technique as well as  
how to compute other transport coefficients with it have been published 
elsewhere~\cite{mueller-plathe:99,mueller-plathe:97a,mueller-plathe:98}.

\section{RESULTS AND DISCUSSION}
\subsection*{Data Comparison with other Work}
\begin{figure}[hb]
\begin{center}
\epsfxsize 12cm \epsfbox {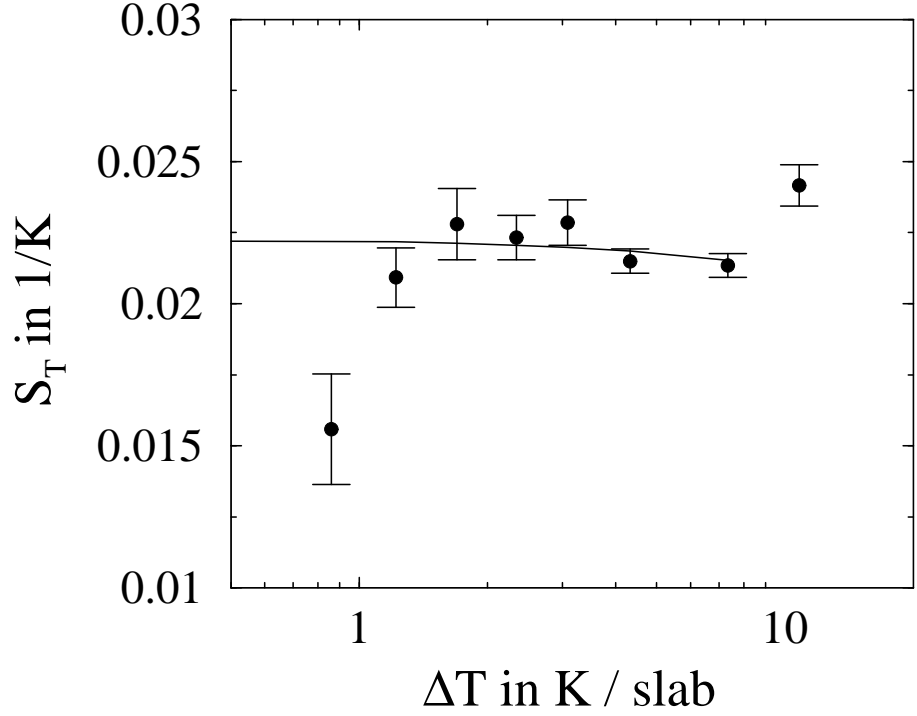}
\end{center}
\caption{Linear response regime and extrapolation to zero temperature gradient.
Model system: 
$m_{1}/m_{2}=4.0$, $\sigma _{1}/\sigma _{2}=1.0$ and $\varepsilon_{1}/
\varepsilon _{2}=1.3$ where species 2 corresponds to Lennard-Jones Argon. 
A straight line was fitted to the data points (with the highest and the lowest
point omitted) in a linear scale. The extrapolation value is 
$S_{T}=22.3\cdot10^{-3}/K$.}
\label{lin_regime}
\end{figure}

Since the Ar-Kr LJ-system was mostly investigated by MD computer simulations
in the past, we made a consistency check of our results with such a system.
Table~\ref{tab:compare} compares our own results with data obtained by several 
other groups, all using identical LJ-parameters for Ar and Kr, respectively 
(Table~\ref{atom_parameters}). For the self diffusion coefficients, the
agreement is excellent. The situation is different concerning the Soret
coefficient. It can be seen, that although the signs for $S_{T}$ are 
identical and all values lie
close to each other, they do not agree with each other within the
statistical uncertainty. However, comparison with the older results is 
somewhat difficult since system sizes were much smaller (128 or 256 particles).
Moreover, we found in our simulations that the value of $S_{T}$ is strongly 
influenced by the specific state point at which
the simulation was run. Minor shifts in the density were e.g.\
sufficient to change the value of $S_{T}$ by more than 30\% 
(Table~\ref{tab:compare}). 
Vogelsang and Hoheisel reported~\cite{vogelsang:88}, that their 
values are systematically too low due to an inaccurate determination of 
the partial enthalpies. Extrapolation to zero field strength 
(MacGowan and Evans~\cite{macgowan:86,macgowan:87}, 
Paolini and Cicotti~\cite{paolini:87}) or zero temperature gradient (as in 
our case) are also sources of error. Figure~\ref{lin_regime} shows that in our
simulations the Soret coefficients can be well extrapolated to zero field. 
The two outermost points deviate from the extrapolation line for two different
reasons. At low temperature gradients (left) there is too much statistical 
uncertainty in the concentration gradient. At high gradient (right) the linear 
response breaks down because of the too large perturbation. Additionally, the 
temperature of the cool region lies then below the freezing temperature.

\subsection*{Systematic Variations of Parameters}
\begin{figure}[hb]
\begin{center}
\epsfxsize 12cm \epsfbox {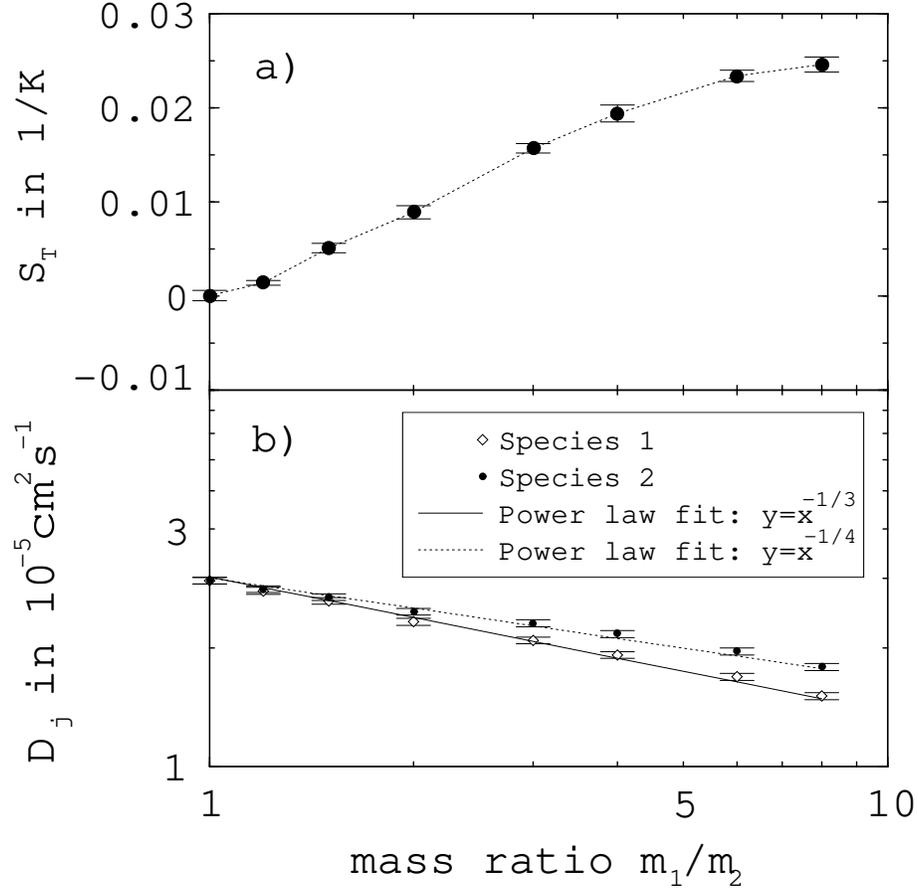}
\end{center}
\caption{(a) Soret coefficient for variations of the mass of species 1
        ($m_{1}>m_{2}$). Species
    2 corresponds to Lennard-Jones Argon. (b) Self diffusion coefficients for the same
    systems.}
\label{F_mass}
\end{figure}

\begin{figure}[hb]
\begin{center}
\epsfxsize 12cm \epsfbox {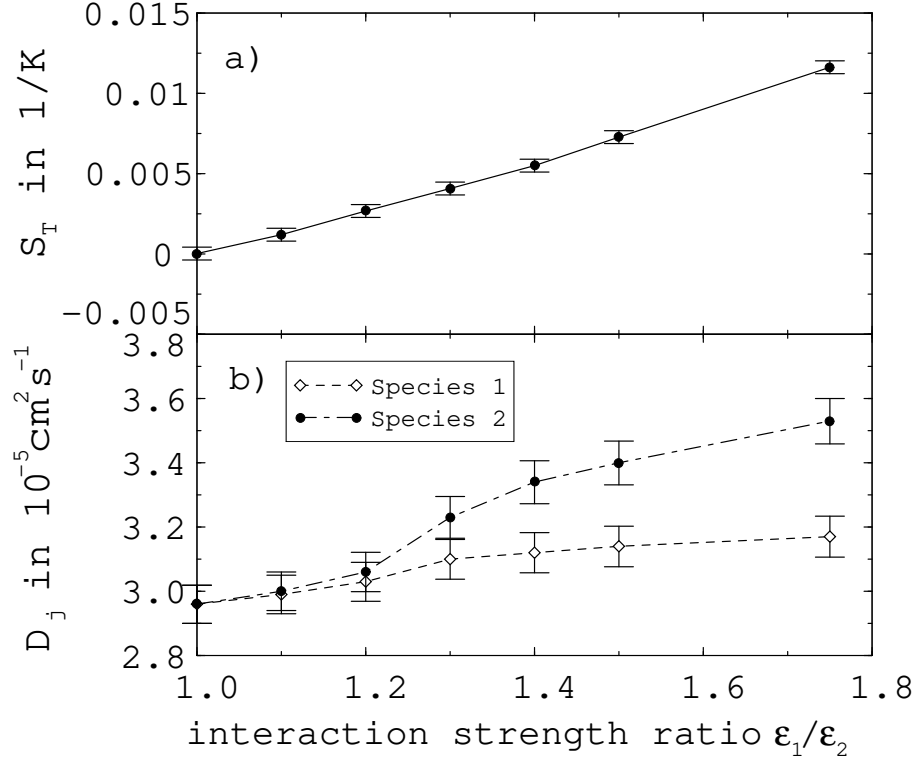}
\end{center}
\caption{(a) Soret coefficient for variations of the potential well depth of 
        species 1. Species 2 corresponds to Lennard-Jones Argon 
        ($\varepsilon_{1}>\varepsilon_{2}$). (b) Self diffusion 
        coefficients for the same systems. In spite of a stronger interaction
        the mobilities increase because the average temperature is raised 
        accordingly, in order to maintain $T^{*}=0.85=const$ 
        (Table \ref{reduced_scales_table}).}
\label{F_epsilon}
\end{figure}

As long as we change only one parameter ratio at a time, the problem is
symmetric in species identity, i.e.\ it is sufficient to consider ratios
larger than $1.0$. We define species 2 as reference species
and identify it with LJ-Argon: $m_{2}=39.95$~amu, $\sigma_{2}=0.3405$~nm and 
$\varepsilon_{2}=1.0$~kJ/mol. When changing the mass ratio while
keeping all other parameters fixed and identical, the positive sign of $S_{T}$
indicates that the
heavier species 1 prefers the cold side of the simulation box whereas the
lighter species 2 favors the hot side (Fig.~\ref{F_mass}). Moreover, $S_{T}$ 
rises monotonically for the whole mass ratio range. 
As a side result 
we found that the behaviour of the self diffusion coefficients could be very 
well fitted by power laws:  $D_{1}\propto (m_{1}/m_{2})^{-0.342}$ 
and $D_{2}\propto (m_{1}/m_{2})^{-0.252}$, which is very close to exponents 
$-\frac{1}{3}$ and $-\frac{1}{4}$. They are also similar to those first found by 
Bearman and Jolly in simulations of Ar-Kr 
mixtures ($-0.3$ and $-0.2$, respectively)~\cite{bearman:81}. For the interaction 
parameter $\varepsilon$ it can be seen that $S_{T}$ is positive for 
$\varepsilon_{1}>\varepsilon_{2}$ (Fig.~\ref{F_epsilon}), i.e.\ species with 
deeper potential wells prefer the cold side of the simulation box while the 
more weakly bound species accumulate on the hot side. The effect of 
mass and interaction strength variations is
intuitively understandable since both times, the species with the lower 
mobility (i.e.\ lower self diffusion coefficient, 
cf.\ Fig.~\ref{F_mass},~\ref{F_epsilon}) favors the cold area. 

\begin{figure}[hb]
\begin{center}
\epsfxsize 12cm \epsfbox {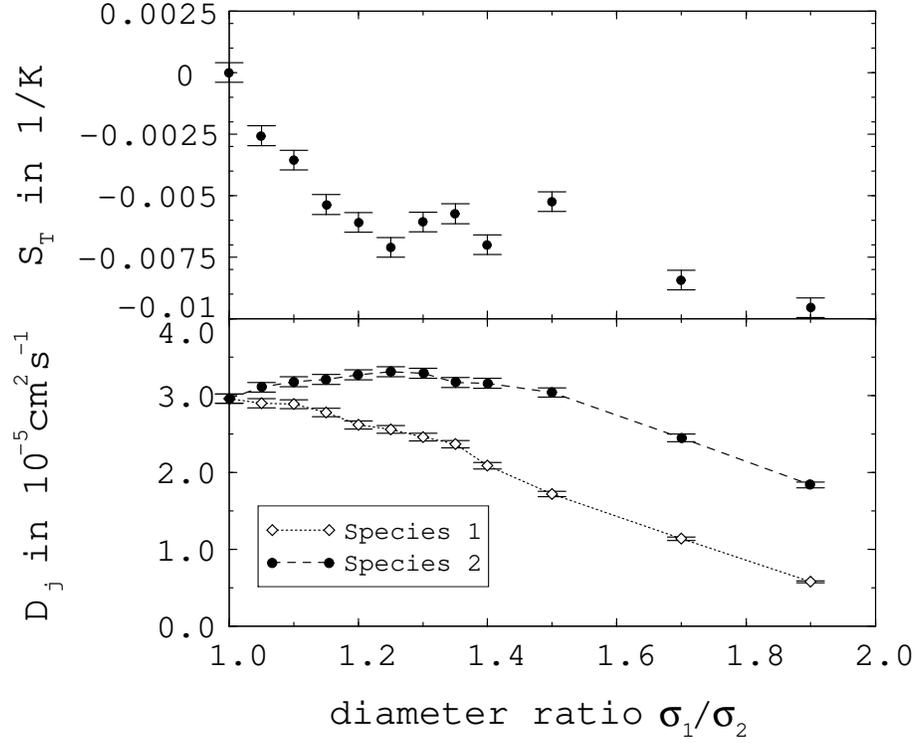}
\end{center}
\caption{(a) Soret coefficient for variations of the diameter of 
        species 1 ($\sigma_{1}>\sigma_{2}$). Species 2 corresponds to 
        Lennard-Jones Argon. (b) Self diffusion coefficients for the same systems.}
\label{F_sigma}
\end{figure}

\begin{figure}[hb]
\begin{center}
\epsfxsize 12cm \epsfbox {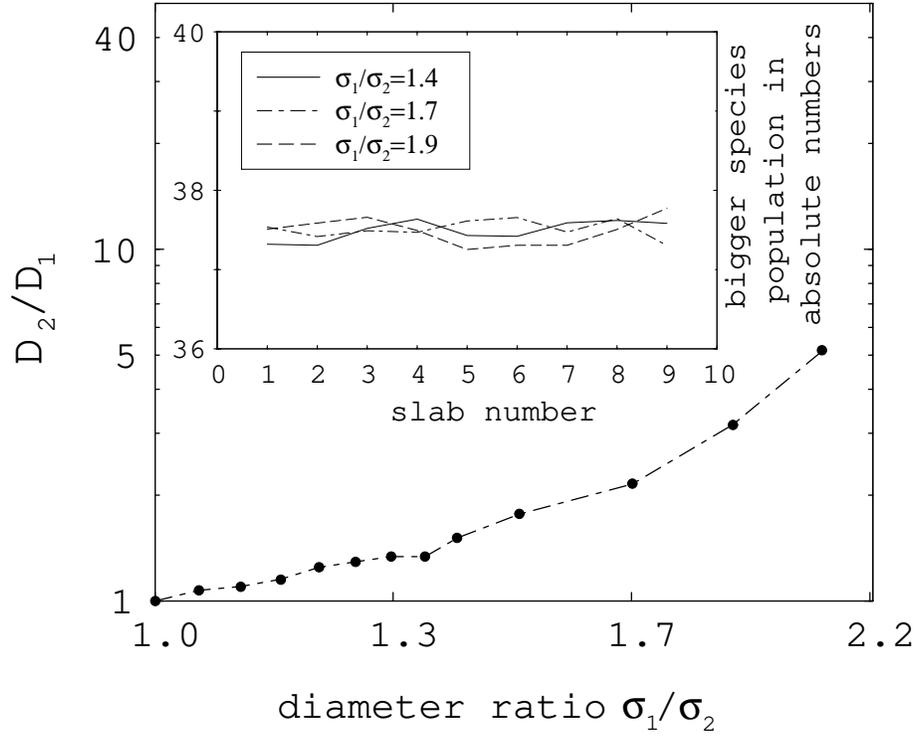}
\end{center}
\caption{Scaling behaviour of the self diffusion coefficient for diameter
        variations. Inset: For large diameter ratios, the slab concentration 
        of the bigger species is nearly independent of the ratio.}
\label{F_sigma_self}
\end{figure}

\begin{figure}[hb]
\begin{center}
\epsfxsize 12cm \epsfbox {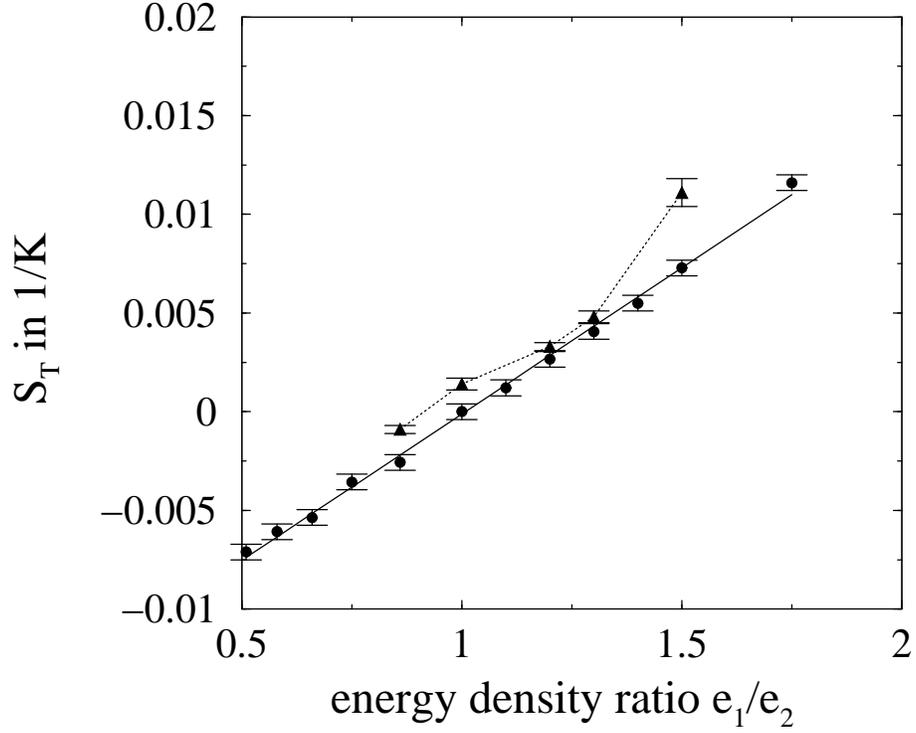}
\end{center}
\caption{The Soret coefficient varies approximately linearly with the ratio of
'cohesive energy densities' $e_{1}/e_{2}>1 (e_{k}=\varepsilon_{k} / \sigma_{k}^{3})$.
Circles: The data for $e_{1}/e_{2}<1$ were calculated from $\sigma_{1} / \sigma_{2}$
variations at constant $\varepsilon_{1} / \varepsilon_{2} = 1$. For the data at  
$e_{1}/e_{2}>1$, $\varepsilon_{1} / \varepsilon_{2}$ was varied at constant
$\sigma_{1} / \sigma_{2}=1$. Triangles: $\varepsilon_{1} / \varepsilon_{2}$ was 
varied at constant $\sigma_{1} / \sigma_{2}=1.1$.
}
\label{F_energy_dens}
\end{figure}

For the atomic diameter parameter $\sigma$, in contrast, we find that
$S_{T}$ is negative for $\sigma_{1}>\sigma_{2}$, i.e.\ smaller particles prefer
the cold side of the simulation box although their mobility is higher compared
to the bigger species (Fig.~\ref{F_sigma}). Furthermore, a 
monotonic behavior 
was only found for $\sigma_{1}/\sigma_{2}<1.25$ (which would already be
large for realistic mixtures). A pressure effect can here be ruled out as the 
pressure did not vary more than 5\%. 
For higher ratios, the $S_{T}$ data points do not show a monotonic
behaviour any more. Here, we also found a large disparity of the two self 
diffusion coefficients. Specifically, the bigger species becomes increasingly 
immobile compared to the smaller species and the absolute number of bigger 
particles per slab is nearly constant (Fig.~\ref{F_sigma_self}). That indicates
that a regime is reached in which the mobility of the larger particles is 
massively hindered by their own space-filling arrangement in the box. 
Therefore, the large particles can no longer exhibit a density gradient.
As a consequence,
the small species remains the only one 
capable to accumulate in the cold region: It is done for entropic reasons 
(the cooler a region, the higher the overall density).
Now the question arises if the trend 'the bigger the species, the more it
tends to accumulate in the hot region' still holds if both species can move 
around unhindered. Therefore, a run with lower density and higher 
temperature ($\sigma_{1}/\sigma_{2}=1.4$, $\rho^{*}=0.6$, $T^{*}=1.15$) was 
performed. The bigger species then builds up a concentration gradient and
accumulates at the hot side with $S_{T}=-6.1\pm0.6$, confirming the trend 
observed for atomic diameter ratios.
However, the high ratio regime is artificial and the above explanation not 
applicable in the realistic area of $\sigma_{1}/\sigma_{2}<1.25$.
There, another argument seems to be more appropriate:
If we combine the LJ parameters $\sigma$ and $\varepsilon$ into the variable
$e=\frac{\varepsilon}{\sigma^{3}}$ we obtain a measure for the stored energy 
per volume, a potential or cohesive energy density. The Soret coefficient varies
linearly with the ratio $e_{1}/e_{2}$ (Fig.~\ref{F_energy_dens}).
The species with higher cohesive energy density accumulates in the cold 
region of the system. 
%

%
%
%
\subsection*{Realistic systems and superposition of different contributions to the Soret coefficient}
\begin{figure}[hb]
\begin{center}
\epsfxsize 12cm \epsfbox {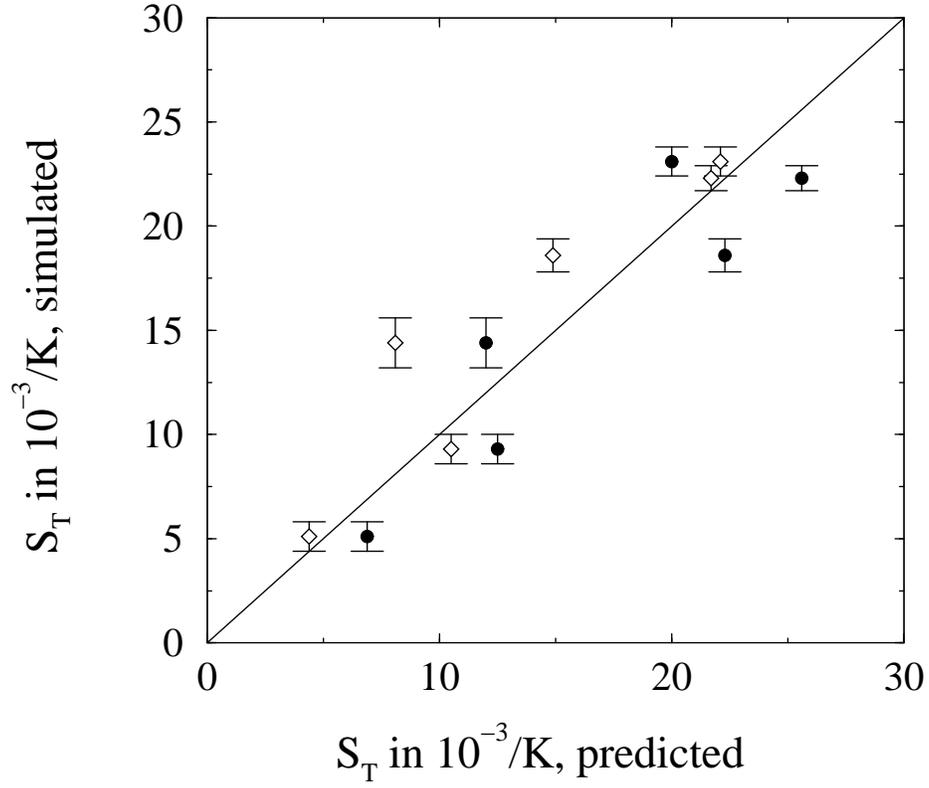}
\end{center}
\caption{ Solid Circles: Correlation of predicted 
        (Eqns.~\ref{90},\ref{91} and~\ref{92}) and simulated Soret coefficients 
        (Table~\ref{tab:05}).
        Open Diamonds: A prediction only using Eqn.~\ref{90} shows that
        the Soret effect is dominated by the mass ratio.}
\label{F_realistic_all}
\end{figure}

We also investigated how more complicated mixtures (species differing in all
parameters) behave under the 
influence of a temperature gradient. Thus, we can check if the 
parameters contribute more or less independently to the Soret coefficient and 
if one of them is dominating. The investigated systems comprised all possible 
liquid binary mixtures of $Ar, Kr, Xe$ and $CH_{4}$ (Table~\ref{tab:05}), 
representing species with strongly deviating parameters. The following empirical 
laws were obtained by low order fits of the independent
parameter variations (Table~\ref{tab:04}):
\begin{eqnarray}
  S_{T} [10^{-3}/K] & = & -0.7\cdot\left(\frac{m_{1}}{m_{2}}\right)^{2}
    +9.5\cdot\left(\frac{m_{1}}{m_{2}}\right)
    -8.8\quad\quad  \mbox{for} \quad m_{1}/m_{2}\leqslant 8.0   
  \label{90} \\ && \nonumber \\
  S_{T} [10^{-3}/K] & = &67.4\cdot\left(\frac{\sigma_{1}}{\sigma_{2}}\right)^{2}
    -179.3\cdot\left(\frac{\sigma_{1}}{\sigma_{2}}\right)
    +111.9\quad\quad  \mbox{for} \quad \sigma_{1}/\sigma_{2}\leqslant 1.25   
  \label{91} \\ &&\nonumber \\
  S_{T} [10^{-3}/K] & = & 4.4\cdot\left(\frac{\varepsilon_{1}}
      {\varepsilon_{2}}\right)^{2}
    +3.5\cdot\left(\frac{\varepsilon_{1}}
      {\varepsilon_{2}}\right)-7.9\quad\quad
         \mbox{for} \quad \varepsilon_{1}/\varepsilon_{2}\leqslant 1.75
  \label{92}
\end{eqnarray}
Using the additivity of the three contributions, 
we predicted the Soret coefficients of the realistic systems
and compared them, using Eqns.~\ref{90},\ref{91} and~\ref{92}, with the simulated 
values (Fig.~\ref{F_realistic_all}). 
The agreement is excellent: The sign of $S_{T}$ is correct for all cases 
and the relative deviation is below $30\%$ for all realistic mixtures.
The results show also clearly that increasing Soret coefficients are mainly
correlated with increasing mass ratios (Fig.~\ref{F_realistic_all}). That is 
because in realistic systems (of unlike noble gases), the mass difference 
between the two species is usually much larger than the deviations in their 
effective diameter or interaction strength. Additionally, the effects of 
$\sigma$ and $\varepsilon$ tend to cancel out each other, since mostly, the 
bigger atom is also the one with the deeper potential well. For this reason it 
is possible to combine the last two equations (\ref{91} and~\ref{92}) into a 
single one which involves the cohesive energy densities. That is done by
fitting a straight line to the composed data shown in Fig.~\ref{F_energy_dens}:
\begin{equation}
  S_{T} [10^{-3}/K] = 14.8\cdot\left(\frac{e_{1}}{e_{2}}\right)-14.8
  \label{93}
\end{equation}
As it can be seen in Table~\ref{tab:05}, the accuracy of the prediction using
Eqns.~\ref{90} and \ref{93} is the same compared to Eqns.~\ref{90}, \ref{91}
and \ref{92} for most data points. Only in the case of the Xe-Ar mixture, the 
predicted values differ significantly. Here, the latter prediction is much 
lower than the first one. That is most probable because the 
individual parameter ratios are large (and, hence, tend to be more inaccurate)
while, due to compensation effects, the composed ratio does not differ much from 
unity compared to the other mixtures. To check if the empirical laws are 
applicable to experimental data, we compared the predicted results
for equimolar mixtures of benzene-chlorobenzene and benzene-nitrobenzene 
(mentioned in the introduction) to the experimental values. 
The sign as well as the order of magnitude of $S_{T}$ could be reproduced:
For benzene-chlorobenzene, $S_{T}(\mbox{pred})=-5.0\cdot10^{-3}$ 
(exp.: $-1.1\cdot10^{-3}$) and for benzene-nitrobenzene 
$S_{T}(\mbox{pred})=-12.6\cdot10^{-3}$  (exp.: $-11.0\cdot10^{-3}$).
In view of the simplicity of the model, we consider this as an excellent 
result which motivates us to continue work in this direction.
\section{CONCLUSIONS}

We investigated with computer simulations thermal diffusion processes in 
simple two-component Lennard-Jones model liquids out of equilibrium. 
We observed that for dense liquid mixtures, the size of the 
Ludwig-Soret effect, depends on the ratios of all three model parameters: 
the mass, the atomic diameter and the interaction strength. However, the
combination of the two latter into a single parameter 
$e=\frac{\varepsilon}{\sigma^{3}}$, the potential
energy density of a species, seems to be especially appropriate to explain 
the sign of the Ludwig-Soret effect. 

Parts of our results are in line with earlier investigations, especially by 
Vogelsang and Hoheisel et al.\ \cite{hoheisel:88,vogelsang:87} and Hafskj{\o}ld 
et al.\ \cite{hafskjold:93,kincaid:94}, while others are completely new.
For isotopic systems, the lighter species can 
transport a higher amount of heat through the system if it accumulates in 
the hot region. Moreover, the possibility to superpose the influence of the 
various parameters was investigated on more realistic mixtures of LJ noble gases. 
For most of these systems, the mass ratio is the 
dominating factor that determines sign and value of $S_{T}$. 
This trend is for realistic mixtures augmented by the fact, 
that the effects of $\sigma$ and $\varepsilon$ tend to cancel out
each other, since bigger species often also are the ones with
the deeper potential depth. Further investigations will compare our 
simulation results with experimental data. 

\section*{Acknowledgements}
We gratefully acknowledge Burkhard D\"unweg, Simone Wiegand and Werner
K\"ohler for fruitful discussions.

\clearpage

\listoftables

\newpage


\begin{table}[htbp]
  \begin{center}
    \begin{tabular}{ll}
\text{\textsl{Quantity}} & \text{\textsl{Reduced units}} \\ \hline \\
\text{Mass} & $m_{j}^{*}=\frac{m_{j}}{m_{Ar}}$ \\[10pt]
\text{Length} & $r^{*}=\frac{r}{\sigma _{Ar}}$ \\[10pt]
\text{Energy} & $E^{*}=\frac{E}{\varepsilon _{Ar}}$ \\[10pt]
\text{Time} & $t^{*}=\frac{t}{\sigma _{Ar}}\sqrt{\frac{\varepsilon _{Ar}}{m_{Ar}}}$ \\[10pt]
\text{Diffusion coefficient} & $D^{*}=\frac{D}{\sigma _{Ar}}\sqrt{\frac{m_{Ar}}{\varepsilon _{Ar}}}$\\[10pt]
\text{Soret coefficient} & $S_{T}^{*}=S_{T}\frac{\varepsilon_{Ar}}{k_{B}}$ \\[10pt]
\text{Pressure} & $p^{*}=p\frac{\sigma _{Ar}^{3}}{\varepsilon_{Ar}}$ \\[10pt]
\text{Density} & $\rho ^{*}=\frac{N\sigma_{12}^{3}}{V}$ \\[10pt]
\text{Temperature} & $T^{*}=\frac{k_{B}T}{\varepsilon _{12}}$ \\[18pt]
    \end{tabular}
    \caption{Reduced units for physical quantities. LJ-Argon was taken to be the
         reference system (Table~\ref{atom_parameters}). The mixed quantities 
        $\sigma_{12}$ 
        and $\varepsilon _{12}$ are derived by the Lorentz-Berthelot rules. For 
        density and temperature they were used to conserve the state point in
        the phase diagram of the mixed system.}
    \label{reduced_scales_table}
  \end{center}
\end{table}

\clearpage
\begin{table}[tbph]
  \begin{center}
    \begin{tabular}{|c|c|c|c|c|c|c|c|}
      & & & & & & &  \\[-10pt] 

      Ref. & Method$^{1}$ & Number & $\rho^{*}$ & $T^{*}$ & $S_{T}$ & $D_{Ar}$ & $D_{Kr}$\\[3pt]

      &  & of atoms  &  &  & $[10^{-3}K^{-1}]$ & $[10^{-5}cm^{2}s^{-1}]$ & $[10^{-5}cm^{2}s^{-1}]$\\
      &  & &  &  & & & \\[-10pt] \hline\hline
      ME$^{2}$ & NEMD & 108,256 & 0.7902 & 0.805 & 11.3 &&\\ \hline
      PC$^{3}$ & NEMD & 108,256 & 0.7902 & 0.824 & 16.2$\pm$2.0 &&\\ \hline
      VH$^{4}$ & EMD & 108,256 & 0.803 & 0.81 & 9.1$\pm$2.0 &&\\ 
      &&& 0.79 & 0.80 &&2.97$\pm$0.08 & 2.60$\pm$0.08\\ \hline
      SH$^{5}$\ & EMD & 256 & 0.79 & 0.81 && 2.97$\pm$0.08 & 2.44$\pm$0.08\\ \hline
      this work & NEMD & 1500 & 0.797 & 0.805 & 10.5$\pm$1.3 & 2.98$\pm$0.05 & 2.48$\pm$0.05\\ \hline
      this work & NEMD & 1500 & 0.81 & 0.85 & 14.4$\pm$1.2 & 2.97$\pm$0.05 & 2.47$\pm$0.05\\ 
    \end{tabular}
  \end{center}
  \caption{Comparison of calculated values for the Soret and self diffusion coefficients of a Lennard-Jones Ar-Kr binary mixture.}
\label{tab:compare}
\end{table}
\vskip -18pt
$^{1}$ EMD: Equilibrium Molecular Dynamics, NEMD: Non-Equilibrium Molecular Dynamics\\
$^{2}$ MacGowan and Evans~\cite{macgowan:86}\\
$^{3}$ Paolini and Cicotti~\cite{paolini:87}\\
$^{4}$ Vogelsang and Hoheisel~\cite{vogelsang:88}\\
$^{5}$ Sch\"on and Hoheisel~\cite{schoen:84} 
\vskip -13pt
\clearpage




\begin{table}[htbp]
  \begin{center}
    \begin{tabular}{crrr}
      Atom species & $m[amu]=m^{*}$ & $\sigma[nm]=\sigma^{*}$ & $\varepsilon[kJ/mol]=\varepsilon^{*}$ \\\hline
      $Ar$ & 39.95=1.00 & 0.3405=1.00 & 1.00=1.00 \\
      $Kr$ & 83.80=2.10 & 0.3633=1.07 & 1.39=1.39 \\
      $Xe$ & 131.29=3.29 & 0.3975=1.17 & 1.72=1.72 \\
      $CH_{4}$ & 16.04=0.40 & 0.3740=1.10 & 1.27=1.27 \\[10pt]
    \end{tabular}
    \caption{Lennard-Jones parameters for the atoms and molecules of this
           study. Molecules are treated as single Lennard-Jones atoms.}
    \label{atom_parameters}
  \end{center}
\end{table}

\clearpage
\begin{table}[tbph]
\begin{center}
\begin{tabular}{|c|c|c|c|c|c|c|}
& & & & & & \\[-10pt] 
$m_{1}/m_{2}$ & $\sigma_{1}/\sigma_{2}$ & $\varepsilon_{1}/\varepsilon_{2}$ & 
$S_{T}$ & $D_{1}$ & $D_{2}$ & $p$\\
&&& $[10^{-3}K^{-1}]$ & $[10^{-5}cm^{2}s^{-1}]$ 
& $[10^{-5}cm^{2}s^{-1}]$ & $[10^{7}Pa]$ \\
& & & & & & \\[-10pt] \hline\hline
1.2  & 1.0 & 1.0 &   1.4$\pm$ 0.2 & 2.96 & 2.96 & 4.6 $\pm$ 0.4  \\ \hline
1.5  & 1.0 & 1.0 &   5.0$\pm$ 0.5 & 2.79 & 2.82 & 4.6 $\pm$ 0.3  \\ \hline
2.0  & 1.0 & 1.0 &   8.9$\pm$ 0.7 & 2.63 & 2.69 & 4.7 $\pm$ 0.3  \\ \hline
3.0  & 1.0 & 1.0 &  15.7$\pm$ 0.5 & 2.30 & 2.37 & 4.7 $\pm$ 0.3  \\ \hline
4.0  & 1.0 & 1.0 &  19.4$\pm$ 0.9 & 2.09 & 2.31 & 4.6 $\pm$ 0.3  \\ \hline
6.0  & 1.0 & 1.0 &  23.4$\pm$ 0.6 & 1.69 & 1.96 & 4.6 $\pm$ 0.4  \\ \hline
8.0  & 1.0 & 1.0 &  24.6$\pm$ 0.8 & 1.51 & 1.79 & 4.6 $\pm$ 0.4  \\ \hline \hline
 1.0 & 1.0 & 1.1 &   1.2$\pm$ 0.3 & 2.99 & 3.00 & 4.9 $\pm$ 0.4  \\ \hline
 1.0 & 1.0 & 1.2 &   2.7$\pm$ 0.4 & 3.03 & 3.06 & 5.0 $\pm$ 0.4  \\ \hline
 1.0 & 1.0 & 1.3 &   4.1$\pm$ 0.5 & 3.10 & 3.23 & 5.2 $\pm$ 0.4  \\ \hline
 1.0 & 1.0 & 1.4 &   5.5$\pm$ 0.8 & 3.12 & 3.34 & 5.3 $\pm$ 0.4  \\ \hline
 1.0 & 1.0 & 1.5 &   7.3$\pm$ 0.7 & 3.14 & 3.40 & 5.3 $\pm$ 0.4  \\ \hline
 1.0 & 1.0 & 1.75&  11.6$\pm$ 0.6 & 3.17 & 3.53 & 5.5 $\pm$ 0.4  \\ \hline
 1.0 & 1.0 & 2.0 &  21.0$\pm$ 0.4 & 3.15 & 3.71 & 5.8 $\pm$ 0.5  \\ \hline
 1.0 & 1.0 & 2.25&  30.7$\pm$ 0.9 & 3.13 & 3.87 & 5.9 $\pm$ 0.6  \\ \hline
 1.0 & 1.0 & 2.5 &  41.4$\pm$ 2.6 & 3.05 & 4.00 & 6.0 $\pm$ 0.7  \\ \hline\hline
 1.0 & 1.05& 1.0 &  -2.6$\pm$ 0.5 & 2.90 & 3.11 & 4.1 $\pm$ 0.3  \\ \hline
 1.0 & 1.1 & 1.0 &  -3.6$\pm$ 0.5 & 2.89 & 3.18 & 3.9 $\pm$ 0.3  \\ \hline
 1.0 & 1.15& 1.0 &  -5.4$\pm$ 0.5 & 2.78 & 3.21 & 3.9 $\pm$ 0.3  \\ \hline
 1.0 & 1.2 & 1.0 &  -6.1$\pm$ 0.5 & 2.62 & 3.27 & 3.8 $\pm$ 0.2  \\ \hline
 1.0 & 1.25& 1.0 &  -7.1$\pm$ 0.5 & 2.56 & 3.31 & 3.8 $\pm$ 0.2  \\ \hline
 1.0 & 1.3 & 1.0 &  -6.1$\pm$ 0.6 & 2.46 & 3.29 & 4.0 $\pm$ 0.2  \\ \hline
 1.0 & 1.35& 1.0 &  -5.7$\pm$ 0.6 & 2.37 & 3.17 & 4.2 $\pm$ 0.2  \\ \hline
 1.0 & 1.4 & 1.0 &  -7.0$\pm$ 0.5 & 2.09 & 3.16 & 4.4 $\pm$ 0.2  \\ \hline
 1.0 & 1.5 & 1.0 &  -5.2$\pm$ 0.5 & 1.72 & 3.04 & 5.0 $\pm$ 0.2  \\ \hline
 1.0 & 1.7 & 1.0 &  -8.4$\pm$ 0.7 & 1.14 & 2.45 & 6.2 $\pm$ 0.1  \\ \hline
 1.0 & 1.9 & 1.0 &  -9.6$\pm$ 1.7 & 0.58 & 1.84 & 7.6 $\pm$ 0.1  \\ 
\end{tabular}
\end{center}
\caption{Thermal Diffusion data: Separate variation of every species 1 parameter
         at the state point $\bar{T}^{*}=0.85$, $\bar{\rho}^{*}=0.81$. 
        Species 2 corresponds to Lennard-Jones Argon. The error is estimated 
        by linear regression (in case of $S_{T}$ of the temperature and 
        mole fraction profiles).
        For the self diffusion coefficients, the error is generally $\pm 2\%$.}
\label{tab:04}
\end{table}

\clearpage
\begin{table}[bph]
\begin{center}
\begin{tabular}{|c|c|c|c|c|c|c|c|c|c|}
& & & & & & & & &\\[-10pt] 
&&&&& simulated & predicted$^{1}$& predicted$^{2}$
& absolute$^{1}$ & relative$^{1}$ \\
Model &$m_{1}/m_{2}$ & $\sigma_{1}/\sigma_{2}$ &
$\varepsilon_{1}/\varepsilon_{2}$ & $e_{1}/e_{2}$ & $S_{T}$ & $S_{T}$ & $S_{T} $ & deviation & deviation \\ 
system&&&&& $[10^{-3}K^{-1}]$ & $[10^{-3}K^{-1}]$ & $[10^{-3}K^{-1}]$ & $[10^{-3}K^{-1}]$ & $[\%]$\\
& & & & & & & & & \\[-10pt] \hline\hline
$Ar_{\mbox{\tiny Iso}}-Ar$ & 1.50 & 1.00 & 1.00 & 1.00 & 5.0$\pm$0.5 & 3.9 & 3.9 &1.1& 22.0 \\ \hline
$Xe-Kr$     & 1.57 & 1.03 & 1.23 & 1.13 & 5.1$\pm$0.7  & 6.2 & 6.3    & -1.1 &-21.6\\ \hline
$Kr-Ar$     & 2.1  & 1.07 & 1.39 & 1.13 & 14.4$\pm$1.2 & 10.8 & 10.1  & 3.6  & 25.0\\ \hline
$Ar-CH_{4}$ & 2.49 & 0.91 & 0.79 & 1.05 &  9.3$\pm$0.7 & 10.7 & 11.2  &-1.4  &-15.0\\ \hline
$Xe-Ar$     & 3.29 & 1.17 & 1.72 & 1.07 & 18.6$\pm$0.8 & 20.4 & 16.0  &-1.8  &-9.7\\ \hline
$Kr-CH_{4}$ & 5.22 & 0.97 & 1.10 & 1.21 & 22.3$\pm$0.6 & 24.3 & 24.8  &-2.0  &-9.0\\ \hline
$Xe-CH_{4}$ & 8.17 & 1.06 & 1.35 & 1.13 & 23.1$\pm$0.7 & 24.5 & 24.1  &-1.4  &-6.1\\ \hline
$Ar_{\mbox{\tiny Iso}}-Ar$ & 8.00 & 1.00 & 1.00 & 1.00 & 24.6$\pm$0.8& 22.4 & 22.4 &2.2&8.9 \\
\end{tabular}
\end{center}
\caption{Soret coefficients for several simple atomic systems
  at the state point $\bar{T}^{*}=0.85$, $\bar{\rho}^{*}=0.81$. The run 
  length was $4.0\cdot 10^{6}$ steps. Additionally, the data for two isotopic
  systems is presented. A comparison indicates, that $S_{T}$ is chiefly
  governed by the mass ratio. The error is estimated 
  by error propagation from the errors of the slopes of temperature and mole
  fraction profiles as obtained in the linear regression.}
\label{tab:05}
\end{table}
\vskip -13pt
  $^{1}$ using Eqns.~\ref{90},\ref{91} and~\ref{92} \\
  $^{2}$ using Eqns.~\ref{90} and \ref{93}
%
%
%
%
%
%
\clearpage

\listoffigures

\newpage

\clearpage

\clearpage

\clearpage

\clearpage


\end{document}